\documentclass[12pt]{article}

\usepackage{latexsym}

\usepackage{graphicx}

\begin{document}


\title{Einstein-Born-Infeld-dilaton black holes in non-asymptotically flat spacetimes }

\author{
     Stoytcho S. Yazadjiev \thanks{E-mail: yazad@phys.uni-sofia.bg}\\
{\footnotesize  Department of Theoretical Physics,
                Faculty of Physics, Sofia University,}\\
{\footnotesize  5 James Bourchier Boulevard, Sofia~1164, Bulgaria }\\
}

\date{}

\maketitle

\begin{abstract}
We derive exact magnetically charged, static and spherically symmetric black hole solutions of
the four-dimensional Einstein-Born-Infeld-dilaton gravity.
These solutions are neither asymptotically flat nor (anti)-de Sitter.
The properties of the solutions are discussed. It is shown that the black holes
are stable against linear
radial perturbations.
\end{abstract}


\sloppy

\section{Introduction}

The nonlinear electrodynamics was first introduced by Born and Infeld in 1934 to obtain
finite energy density model for the electron \cite{BI}. In recent years nonlinear
electrodynamics models are attracting much interest, too. The reason is that the nonlinear
electrodynamics arises naturally in open strings and $D$-branes \cite{FT}-\cite{L}.
Nonlinear electrodynamics models coupled to gravity have been discussed  in different
aspects (see for example \cite{MD}-\cite{CPW} and references therein).

In the present work we consider stringy Einstein-Born-Infeld-dilaton (EBId) gravity described
by the action \cite{MRT}-\cite{AT}

\begin{equation}\label{EBIDA}
{\cal S} = \int d^4x\sqrt{-g}\left[{\cal R} -2g^{\mu\nu}\partial_{\mu}\varphi\partial_{\nu}\varphi
+ L_{BI}\right]
\end{equation}

where ${\cal R}$ is Ricci scalar curvature with respect to the spacetime metric $g_{\mu\nu}$ and $\varphi$ is the dilaton field. The Born-Infeld (BI) part of the action is given by

\begin{equation}
L_{BI} = 4be^{2\gamma\varphi} \left[1 -\sqrt{1 + {e^{-4\gamma\varphi} \over 2b}F^{2}
- {e^{-8\gamma\varphi}\over 16b^2} \left(F\star F \right)^2   } \right].
\end{equation}

Here $\star F$ is the dual to the Maxwell tensor and $\gamma$ ($\gamma\ne 0$) is the dilaton
coupling constant. In the context of the string theory, the (BI) parameter $b$ is related to
the string tension $\alpha$ by $b=1/2\pi\alpha$. It should be noted that the EBId action
does not posses an electric-magnetic duality. That is why one should expect that
the electrically and magnetically charged solutions
will be different. Note that in the $b\to \infty$ limit the
action (\ref{EBIDA}) reduces to Einstein-Maxwell-dilaton
one.

Unfortunately, the field equations yielded by the action (\ref{EBIDA}) are too complicated and
there are no exact
analytical solutions in four or more dimensions. Exact solutions to the EBId equations
are known only in three dimensions \cite{YI}. These solutions are non-asymptotically
flat and describe three-dimensional black holes.

In the present paper we derive exact non-asymptotically flat and non-(A)dS  black hole
solutions to the four-dimensional EBId gravity. Such type solutions, which are
non-charged or within the framework of linear electrodynamics have attracted much
interest in recent years \cite{MW1}-\cite{Y}.

\section{Basic equations}

Here we consider only magnetically charged case for which $F\star F=0$ and that is
why we may restrict ourselves
to the truncated BI Lagrangian

\begin{equation}
L_{BI} = 4be^{2\gamma\varphi} \left[1 -\sqrt{1 + {e^{-4\gamma\varphi} \over 2b}F^{2} } \right].
\end{equation}

It is also convenient to set

\begin{equation}
L_{BI} = 4be^{2\gamma\varphi} {\cal L}_{BI}(X)
\end{equation}

where

\begin{eqnarray}
{\cal L}_{BI}(X) &=& 1 - \sqrt{1 + X}, \\
X &=& {e^{-4\gamma\varphi} \over 2b} F^{2}.
\end{eqnarray}

The action (\ref{EBIDA}) then yields  the following field equations

\begin{eqnarray}\label{FE}
{\cal R}_{\mu\nu} = 2\partial_{\mu}\varphi\partial_{\nu}\varphi
- 4e^{-2\gamma\varphi}\partial_{X}{\cal L}_{BI}(X) F_{\mu\beta} F_{\nu}^{\beta}
+ 2b e^{2\gamma\varphi} \left[2X\partial_{X}{\cal L}_{BI}(X)
- {\cal L}_{BI}(X) \right]g_{\mu\nu} ,\\
\nabla_{\mu}\left[e^{-2\gamma\varphi} \partial_{X}{\cal L}_{BI}(X) F^{\mu\nu}\right]= 0,\\
\nabla_{\mu}\nabla^{\mu}\varphi
=  2b\gamma e^{2\gamma\varphi}\left[2X\partial_{X}{\cal L}_{BI}(X) - {\cal L}_{BI}(X)\right],
\end{eqnarray}

where $\nabla_{\mu}$ is the covariant derivative with respect to the spacetime metric $g_{\mu\nu}$.

The metric of the static and spherically symmetric spacetime can be written in the form

\begin{equation}
ds^2 = - \lambda(r)dt^2 + {dr^2\over  \lambda(r)}
+ R^2(r)\left(d\theta^2 + \sin^2\theta d\phi^2 \right).
\end{equation}

The electromagnetic  field is assumed to have the following pure magnetic form

\begin{equation} \label{MFMF}
F = P\sin\theta d\theta \wedge d\phi
\end{equation}

where $P$ is the magnetic charge. Respectively, we obtain for $X$:

\begin{equation}
X = e^{-4\gamma\varphi} {P^2\over bR^4} .
\end{equation}

The field equations reduce to the following system of coupled ordinary differential equations

\begin{eqnarray}\label{ODE}
{1\over 2} {d\over dr }\left(R^2 {d\lambda\over dr } \right) &=&
-2be^{2\gamma\varphi} \left[2X\partial_{X}{\cal L}_{BI}(X) - {\cal L}_{BI}(X)\right]R^2 ,
\nonumber \\
-{1\over R}{d^2R\over dr^2} &=& \left({d\varphi\over dr }\right)^2 ,\\
1 - {1\over 2} {d\over dr }\left(\lambda {dR^2\over dr } \right) &=&
- 2be^{2\gamma\varphi}{\cal L}_{BI}(X) R^2 ,\nonumber \\
{d\over dr }\left(R^2 \lambda {d\varphi\over dr } \right) &=&
2\gamma be^{2\gamma\varphi} \left[2X\partial_{X}{\cal L}_{BI}(X)
- {\cal L}_{BI}(X)\right]R^2 . \nonumber
\end{eqnarray}

\section{Black holes with string coupling constant $\gamma=1$ }

The case $\gamma=1$ is predicted from the (super)string theory.
In order to solve the field equations we make the ansatz

\begin{equation}\label{ansatz1}
R^2(r)e^{2\varphi} =r^2_{0}
\end{equation}

where $r_{0}>0$ is a constant. The second equation of (\ref{ODE}) then gives

\begin{equation}
R^2(r) = R^2_{0} \left({r- C\over r_{0} }\right)
\end{equation}

where $C$ and $R_{0}>0$ are constants.

The consistency condition for the third and the fourth equation of (\ref{ODE})
gives the following algebraic equation for $r_{0}$:

\begin{equation}
{1\over r^2_{0}} + 4bX\partial_{X}{\cal L}_{BI}(X) = 0
\end{equation}

with $X = {P^2\over br^4_{0} }$. Solving this equation with respect to $r_{0}$ we obtain

\begin{equation}
r^2_{0} = 2P^2 \sqrt{1 -{P^2_{crit}\over P^2}}
\end{equation}

where $P^2_{crit}=1/4b$. Therefore the magnetic charge must satisfy the inequality

\begin{equation}
P^2 > P^2_{crit}.
\end{equation}

The existence of  critical value $P_{crit}$ for the magnetic charge is a pure nonlinear
effect which disappears in
the limit to linear electrodynamics when $b\to \infty$.

Finally, for the metric function $\lambda(r)$ we find

\begin{equation}
\lambda(r) = A {r_{0}(r -r_{+})\over R^2_{0} }
\end{equation}

where

\begin{equation}
A = 4 b \left[{\cal L}_{BI}(X) - 2 X\partial_{X}{\cal L}_{BI}(X) \right]r^2_{0} >0
\end{equation}

and $r_{+}$ is a constant. Below we discuss the physical properties of the solution and,
without loss of generality, we set $C=0$.

The solution is not asymptotically flat and in order to define its  mass we use the
so-called quasilocal formalism \cite{BY}. The  quasilocal  mass is given by

\begin{equation}\label{QLM}
{\cal M}(r) = {1\over 2} {dR^2(r)\over dr} \lambda^{1/2}(r)\left[\lambda_{0}^{1/2}(r)
-\lambda^{1/2}(r) \right]
\end{equation}

where $\lambda_{0}(r)$ is an arbitrary non-negative function which determines the zero
of the energy for a background spacetime. If no cosmological horizon is present,
the large $r$ limit of (\ref{QLM}) determines the asymptotic  mass $M$.
In our case the natural choice is $\lambda_{0}(r) = A r_{0}r/R^2_{0}$ and we find

\begin{equation}
M = {1\over 4}A r_{+}.
\end{equation}
We first consider solutions with positive mass, $M>0$.  The Kretschmann scalar is

\begin{equation}\label{KS}
{\cal K} = {\cal R}_{\mu\nu\alpha\beta} {\cal R}^{\mu\nu\alpha\beta}
= 4 {\cal K}^2_{1} + 8 {\cal K}^2_{2}
+ 8 {\cal K}^2_{3} + 4 {\cal K}^2_{4}
\end{equation}

where

\begin{eqnarray}
{\cal K}_{1} &=& {\cal R}^{01}{}_{01} =0
,\nonumber \\
{\cal K}_{2} &=& {\cal R}^{02}{}_{02} = - {1\over 4} {Ar_{0}\over R^2_{0} r} ,\\
{\cal K}_{3} &=& {\cal R}^{12}{}_{12} =
{1\over 4} {Ar_{+}\over R^2_{0}r_{0}} \left({r_{0}\over r}\right)^2 ,\nonumber \\
{\cal K}_{4} &=& {\cal R}^{23}{}_{23} = {1\over R^2_{0}} \left({r_{0}\over r}\right)^2
\left[{r\over r_{0} } - \left({A\over 4}\right){r - r_{+}\over r_{0} } \right]\nonumber .
\end{eqnarray}

The scalar ${\cal K}$ is singular only for $r=0$ and tends to zero like $1/r^2$ for $r\to \infty$.
The solution has a regular horizon at $r=r_{+}$ which hides a spacelike singularity
located at $r=0$. Note that the dilaton field is regular on the horizon, too.
The spacial infinity is conformally null and the solution describes a black hole
with the same causal structure as the Schwarzschild spacetime. The temperature and
the entropy of the black hole are

\begin{eqnarray}
T &=& {1\over 4\pi} {d\lambda \over dr}(r_{+}) = {1\over 4\pi}{Ar_{0}\over R^2_{0}}, \\
S &=& \pi R^2_{0} {r_{+}\over r_{0} }.
\end{eqnarray}

In order to write the first law we should take into account that the
constants $R_{0}$,$A$ and $r_{0}$
are, in fact, related to the background, not to the black hole. Then we find

\begin{equation}
dM = TdS
\end{equation}

where the variation is with respect to $r_{+}$ only.

The solution with zero mass ($M=0$) is singular with a null singularity at $r=0$. The case $M<0$ corresponds to
naked timelike singularities located at $r=0$.

\section{Black holes with general dilaton coupling }

Here we consider solutions with general dilaton coupling constant $\gamma$. As for the
particular case $\gamma=1$ we make the ansatz

\begin{equation}\label{ansatz2}
R^{2}(r)e^{2\gamma\varphi} = r^2_{0}
\end{equation}

where $r_{0}>0$ is a constant.

Substituting in the equations (\ref{ODE}) we obtain the following algebraic equation
for $r_{0}$

\begin{equation}
{1\over 2r^2_{0}} = - (1-\gamma^2) b{\cal L}_{BI}(X) - 2\gamma^2 b X\partial_{X}{\cal L}_{BI}(X)
\end{equation}

where $X = {P^2\over br^4_{0}}$. In more explicit form the algebraic equation is

\begin{equation}\label{AE}
F(z)=1 - (1-\gamma^2)z - \gamma^2 z^2 - {P_{crit}\over |P|}\sqrt{1 -z^2}=0
\end{equation}

where

\begin{eqnarray}
z &=& {1\over \sqrt{1+ X}} ,\\
r^2_{0} &=& 2|PP_{crit}| {z \over \sqrt{1 -z^2} }.
\end{eqnarray}

For sufficiently small $\varepsilon >0$  we have $F(1-\varepsilon)<0$ and, therefore,
a sufficient (but not necessary) condition for the algebraic eq. (\ref{AE}) to have
roots in the interval $0<z<1$  is $F(0)= 1 - {P_{crit}\over|P|}>0$. In general,
the function $F(z)$ can have zeros also for $P_{crit}>|P|$ and, in some cases,
more than one zero as it is shown in the Figure 1 for $\gamma=\sqrt{3}$. Let us note that
the different zeros of $F(z)$, in fact, correspond to different backgrounds rather to
different black holes, as it is seen from the metric function $\lambda(r)$ presented
below.

For the metric functions we find\footnote{Unimportant constants have been omitted. }

\begin{eqnarray}
R^2(r) &=& R^2_{0} \left({r\over r_{0}} \right)^{{2\gamma^2\over 1 + \gamma^2 }}, \\
\lambda(r) &=&
 A \left({r\over r_{0}} \right)^{{1 - \gamma^2\over 1 + \gamma^2 }}
{r_{0}(r - r_{+})\over R^2_{0} },
\end{eqnarray}

where $R_{0}>0$ and $r_{+}$ are constants and $A$ is given by

\begin{equation}
A = 2(1+ \gamma^2)b \left[{\cal L}_{BI}(X) - 2 X\partial_{X}{\cal L}_{BI}(X) \right]r^2_{0} >0 .
\end{equation}

The natural background is given by the metric function
\begin{equation}
\lambda_{0}(r)= A \left({r\over r_{0}} \right)^{{1 - \gamma^2\over 1 + \gamma^2}}
{r r_{0}\over R^2_{0} }.
\end{equation}

For the asymptotic mass we then obtain

\begin{equation}
M = {\gamma^2 A \over 2(1 + \gamma^2)} r_{+} .
\end{equation}

We first consider solutions with positive mass, $M>0$.

The Kretschmann scalar is given by (\ref{KS}) where

\begin{eqnarray}
{\cal K}_{1} &=& {\cal R}^{01}{}_{01} = {1 -\gamma^2\over 1 + \gamma^2 }
{A\over R^2_{0} } \left({r_{0}\over r } \right)^{{1+ 3\gamma^2\over 1 + \gamma^2}} \left({\gamma^2\over 1+ \gamma^2 } {r-r_{+}\over r_{0} } - {r\over r_{0} } \right)
,\nonumber \\
{\cal K}_{2} &=& {\cal R}^{02}{}_{02} =
- {1\over 2}{\gamma^2\over 1
+ \gamma^2 } {A\over R^2_{0} } \left({r_{0}\over r } \right)^{{1+ 3\gamma^2\over 1
+ \gamma^2}} \left({1-\gamma^2\over 1+ \gamma^2 } {r-r_{+}\over r_{0} } + {r\over r_{0} } \right) ,\\
{\cal K}_{3} &=& {\cal R}^{12}{}_{12} ={1\over 2}{\gamma^2\over 1 + \gamma^2 }
 {A\over R^2_{0} } \left({r_{0}\over r } \right)^{{1+ 3\gamma^2\over 1 + \gamma^2}}
{r_{+}\over r_{0} } ,\nonumber \\
{\cal K}_{4} &=& {\cal R}^{23}{}_{23} =
 - {1\over R^2_{0} } \left({r_{0}\over r } \right)^{{1+ 3\gamma^2\over 1 + \gamma^2}}
\left[{\gamma^4 A\over  (1+ \gamma^2)^2} {r- r_{+}\over r_{0} } - {r\over r_{0}} \right] \nonumber .
\end{eqnarray}

The Kretschmann scalar is divergent only for $r=0$ and tends to zero
like $r^{-{4\gamma^2\over 1 + \gamma^2}} $
for $r\to \infty$. The solution has a regular horizon at $r=r_{+}$ hiding a
spacelike singularity located at
$r=0$. For $\gamma^2<1$, spacial infinity is conformally timelike and the causal structure
is similar to that of the static BTZ black hole spacetime \cite{BTZ}. For $\gamma^2 \ge 1$,
spacial infinity is conformally null  and the causal structure is just the same
as for the Schwarzschild spacetime.
The temperature and the entropy of the black holes are given by

\begin{eqnarray}
T &=& {Ar_{0}\over 4\pi R^2_{0} } \left({r_{+}\over r_{0}} \right)^{{1-\gamma^2\over 1
+ \gamma^2 }}, \\
S &=& \pi R^2_{0} \left({r_{+}\over r_{0}}  \right)^{{2\gamma^2\over 1 + \gamma^2 }}.
\end{eqnarray}

The first law is written in the form

\begin{equation}
dM = T dS
\end{equation}
where the variation is with respect to $r_{+}$ only.

The solutions with zero mass are singular with null singularities.
The case with negative mass is also
singular with null singularities for $\gamma^2\le 1$ and timelike singularities for $\gamma^2>1$.

\section{Linear stability}

The stability of the black holes is an important question from physical point of view.
It is well known that
there are many black holes solutions which are unstable. Here we show that our black hole
solutions are stable
against linear radial perturbations. In order to discuss the stability we take the spacetime
metric in the form

\begin{equation}\label{TDM}
ds^2 = - e^{\Gamma}dt^2 + e^{\chi}dr^2 + e^{\beta}(d\theta^2 + \sin^2\theta d\phi^2)
\end{equation}

where the functions $\Gamma$, $\chi$ and $\beta$ depend on $r$ and $t$. We assume that the
metric functions and the
dilaton are small perturbations of the static background

\begin{eqnarray}
\Gamma(r,t) = \ln \lambda(r) + \delta\Gamma(r,t) ,\,\,\,\,\,\,\,\, \chi(r,t) = -\ln \lambda(r) + \delta\chi(r,t) , \nonumber \\
\beta(r,t) = 2\ln R(r) + \delta\beta(r,t) ,\,\,\,\,\,\,\, \varphi(r,t) = \varphi(r) + \delta\varphi(r,t) .
\end{eqnarray}

The convenient gauge is $\delta\beta(r,t)=0$ (i.e. $e^{\beta}=R^2(r)$). The electromagnetic
field is given by (\ref{MFMF}) which solves the electromagnetic equations for the time
dependent metric (\ref{TDM}), too.

The linearized equations for ${\cal R}_{10}$ and ${\cal R}_{22}$ give

\begin{eqnarray}\label{PR01}
{1\over 2}\partial_{r}\beta(r)\partial_{t} \delta\chi =
2 \partial_{r}\varphi(r) \partial_{t}\delta\varphi,
\end{eqnarray}

\begin{eqnarray}\label{PR22}
\left[1 - {\cal R}_{22}^{(0)} \right] \delta\chi -
{1\over 4}\lambda(r)e^{\beta(r)}  \partial_{r}\beta(r) \left(\partial_{r}\delta\Gamma
 - \partial_{r}\delta\chi\right) \nonumber \\ =
4abe^{2\gamma\varphi(r)}\left[2X\partial_{X}{\cal L}_{BI}(X)
- {\cal L}_{BI}(X)\right] e^{\beta}\delta\varphi,
\end{eqnarray}

where ${\cal R}_{\mu\nu}^{(0)}$ is the Ricci tensor with respect to the static background.
The linearized equation for the dilaton is

\begin{eqnarray}\label{DPER}
\nabla_{\mu}^{(0)}{\nabla^{(0)}}^{\mu}\delta\varphi
- \left[\nabla_{\mu}^{(0)}{\nabla^{(0)}}^{\mu}\varphi(r)\right]\delta\chi
+ {1\over 2}\lambda(r)\partial_{r}\varphi(r) \left(\partial_{r}\delta\Gamma
- \partial_{r}\delta\chi \right) = \nonumber \\
- 4\gamma^2 be^{2\gamma\varphi(r)} \left[{\cal L}_{BI}(X)
+ 4X^2\partial^2_{X}{\cal L}_{BI}(X) \right]\delta\varphi
\end{eqnarray}

where $\nabla_{\mu}^{(0)}$ is the coderivative operator with respect to the static background.

Integrating the equation (\ref{PR01}) we obtain

\begin{equation}\label{CCF}
\delta\chi = 4{\partial_{r}\varphi(r)\over \partial_{r}\beta(r)} \delta\varphi =
 - {2\over \gamma}\delta\varphi .
\end{equation}

Eliminating the perturbations $\delta\chi$ and $\delta\Gamma$
between Eqs. (\ref{PR22}), (\ref{CCF}) and (\ref{DPER})
we find

\begin{equation}\label{GPE}
\nabla_{\mu}^{(0)}{\nabla^{(0)}}^{\mu}\delta\varphi - U(r)\delta\varphi = 0
\end{equation}

where

\begin{equation}
U(r) = 4be^{2\gamma\varphi(r)} \left[{\cal L}_{BI}(X) - 2X\partial_{X}{\cal L}_{BI}
- \gamma^2\left[{\cal L}_{BI}(X) + 4X^2\partial^2_{X}{\cal L}_{BI}(X) \right] \right] .
\end{equation}

Taking into account the explicit form of the BI Lagrangian it can be shown that $U(r)>0$.

Introducing the new radial coordinate

\begin{equation}
r_{*} = r_{0}r_{+}\int {dr\over \lambda(r) e^{\beta(r)} } =
 Ar_{0}\ln\left(1 - {r_{+}\over r } \right)
\end{equation}

the Eq.(\ref{GPE}) can be written in the explicit form

\begin{equation}
- {e^{2\beta(r_{*})}\over r^2_{0}r^2_{+}}\partial^2_{t} \delta\varphi
+ {d^2\delta\varphi\over dr^2_{*} }
- U(r_{*})\lambda(r_{*}) {e^{2\beta(r_{*})} \delta\varphi\over r^2_{0}r^2_{+}} = 0
\end{equation}

For growing modes $\delta\varphi(r,t)=\delta\varphi(r)e^{\eta t}$ we find

\begin{equation}
-{d^2\delta\varphi(r_{*})\over dr^2_{*}}
+ {e^{2\beta(r_{*})} \over r^2_{0}r^2_{+}}\left[\eta^2 + U(r_{*})\lambda(r_{*}) \right]
\delta\varphi(r_{*})=0.
\end{equation}

For real $\eta$ the effective potential
$U_{eff}= {e^{2\beta}\over r^2_{0}r^2_{+}}\left[\eta^2 + U\lambda\right]$ is positively defined
for $r>r_{+}$. Therefore, there are no bounded solutions for $\delta\varphi$ and
we conclude that the black holes
are stable against linear radial perturbations.

\section{Conclusion}

In this paper we derived exact, magnetically charged, static and spherically symmetric black hole
solutions to the Einstein-Born-Infeld-dilaton gravity. These solutions are neither
asymptotically flat nor (anti)-de Sitter. Some basic properties of the solutions were discussed.
It was shown that the black holes are stable against
linear radial perturbations. It is worth noting that the black solutions derived in the present
paper are solutions not only of the BI electrodynamics but also of general nonlinear
electrodynamics  described by an arbitrary function
${\cal L}(X)$ provided the corresponding algebraic equations possess roots and the corresponding
algebraic inequalities are satisfied.
In particular, in the case of the linear electrodynamics with ${\cal L}(X)= -{1\over 2}X$the method presented here gives the well-known
non-asymptotically flat and non-(A)dS black hole solutions of Einstein-Maxwell-dilaton (EMd)
gravity \cite{CHM}. Moreover, for  ${\cal L}(X)= -{1\over 2}X$ we have $U(r)>0$ and therefore,
the EMd black holes are stable\footnote{The stability of the EMd black holes in the particular
case $\gamma=1$ was proven in \cite{CGL}. } against linear radial perturbations for
arbitrary dilaton coupling constant $\gamma$.

It is puzzling that the nonlinear electrodynamics equations can be solved for arbitrary function
${\cal L}(X)$. The cause is that we consider the sector where the theory looses a part
of its nonlinearity\footnote{I am grateful to one of the
referees for turning my attention to this point.}.
More precisely, the electromagnetic nonlinearity of the differential equations is
transformed into algebraic nonlinearity. This can be explicitly demonstrated  as follows.
For spherically symmetric magnetic configurations the ansatz (\ref{ansatz1}) and (\ref{ansatz2})
are equivalent to consider $X$ satisfies

\begin{equation}
X=X_{0}=const.
\end{equation}

The first consequence following from this fact is that the nonlinear Maxwell equations (\ref{FE})
become linear since the nonlinear part can be pulled out of the equations. In second place,
redefining the dilaton field as ${\tilde \varphi}=\varphi - \varphi_{0}$ where

\begin{equation}
\varphi_{0} = {1\over 2\gamma}\left.
\ln\left( 2[{\cal L}(X) -2X\partial_{X}{\cal L}(X)]\over X\right)\right|_{X=X_{0}},
\end{equation}

the dilaton equation becomes the one of the linear EMd case. The only
difference appears in the Einstein equations, which reduce to

\begin{equation}
{\cal R}_{\mu\nu}= 2\partial_{\mu}{\tilde \varphi}\partial_{\nu}{\tilde \varphi}
+ 2ke^{-2\gamma{\tilde \varphi}} F_{\mu\beta}F_{\nu}^{\beta}
- {1\over 2}e^{-2\gamma{\tilde \varphi}} F^2g_{\mu\nu}
\end{equation}

where

\begin{equation}\label{NC}
k = \left.{X\partial_{X}{\cal L}(X)\over 2X\partial_{X}{\cal L}(X) - {\cal L}(X)}\right|_{X=X_{0}}.
\end{equation}

In this way we obtained field equations which are linear in the electromagnetic  field.
Note however, although linear in the electromagnetic  field, these equations, in general,
are not the EMd equations since $k\ne 1$ for general nonlinear
electrodynamics. For example, in the case of the Born-Infeld electrodynamics we have $k\ne 1$
for any finite value of $X_{0}$. Summarizing, we have shown that the all information about
the nonlinearity is encoded in the parameter $k$ appearing in the Einstein equations and
the nonlinear algebraic constraint (\ref{NC}).
It can be seen from the exact solutions presented in the previous sections, that the solutions
of the "k-deformed" EMd equations are quite similar to those of the pure EMd equations.
The parameter $k$ influences the background constants in the solutions. In contrast, the nonlinear
constraint (\ref{NC}) gives severe physical restrictions on the black holes charge. These
restrictions algebraically  reflect the nonlinearity of the electromagnetic field. This
can be explicitly seen from the solutions for the Born-Infeld electrodynamics with $\gamma=1$
where the constraint (\ref{NC}) gives the existence condition $P^2>P^2_{crit}$.

\section*{Acknowledgements} I would like to thank A. Donkov for reading the manuscript. This work
was partially supported by the Bulgarian National Science Fund under Grant MU-408.

\begin{figure}
\begin{center}
    \includegraphics[width=6.5cm]{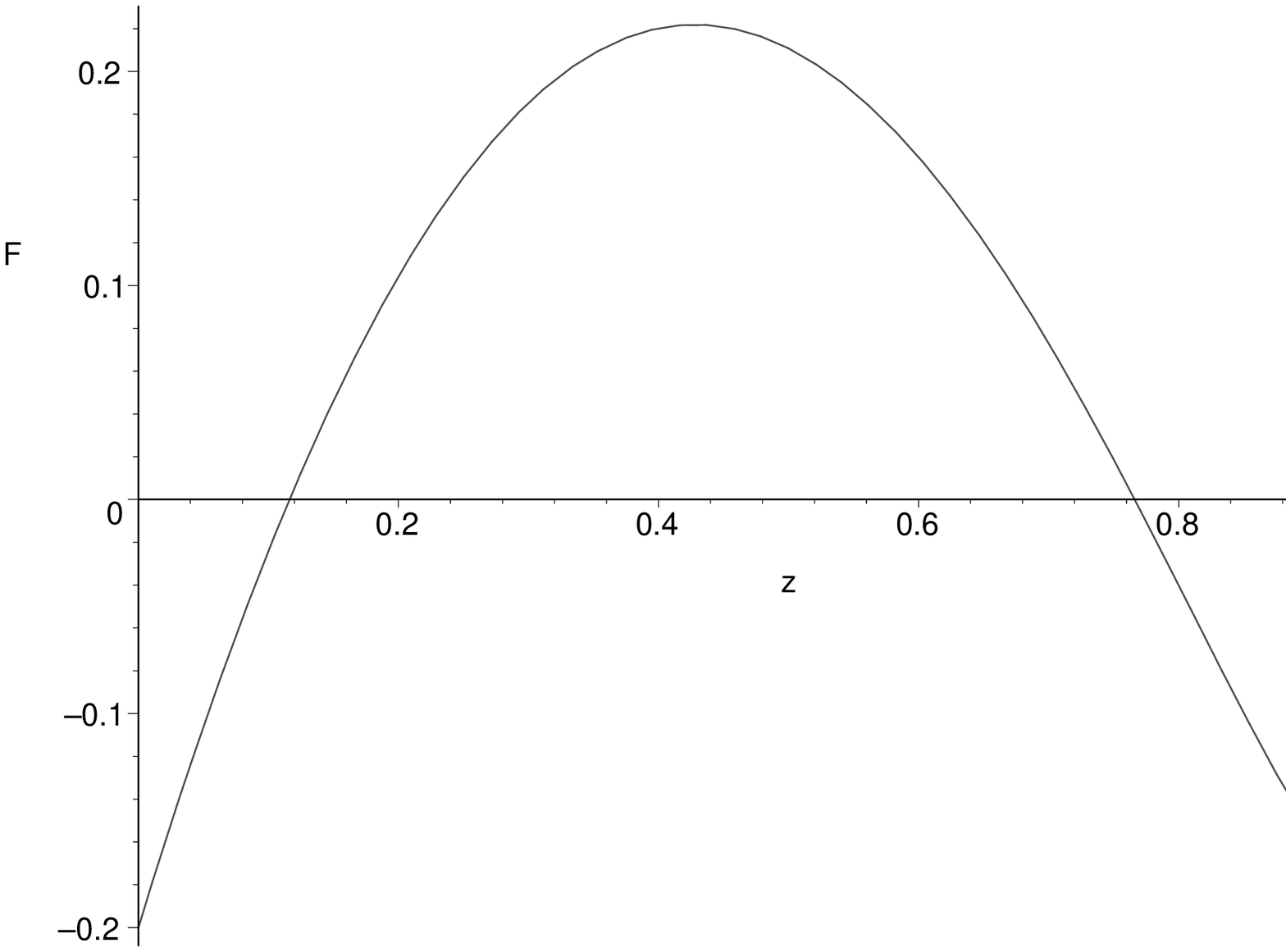}\\
  \caption{Behavior of the function  $F(z)$ for $\gamma=\sqrt{3}$ and ${P_{crit}\over |P|}= 1.2 $ }\label{fig3}
   \end{center}
\end{figure}

\end{document}